\documentclass[preprint,aps ,nofootinbib]{revtex4}
\usepackage{graphicx}
\usepackage{amsmath}
\usepackage{amsfonts}
\usepackage{amssymb}
\usepackage{color}%
\usepackage{dcolumn}

\providecommand{\U}[1]{\protect\rule{.1in}{.1in}}

\newcommand{\f}{\begin{equation}}
\newcommand{\ff}{\end{equation}}
\newcommand{\fa}{\begin{eqnarray}}
\newcommand{\ffa}{\end{eqnarray}}

\begin{document}

\title{Some properties of the holographic fermions in an extremal charged dilatonic black hole}
\author{Jian-Pin Wu}
\email{jianpinwu@mail.bnu.edu.cn}
\affiliation{Department of Physics, Beijing Normal University, Beijing 100875, China}

\begin{abstract}

We study the properties of the Green¡¯s functions of the fermions in the extremal charged dilatonic black
hole proposed by Gubser and Rocha [Phys. Rev. D \textbf{81}, 046001 (2010)]. We find that many properties seem
to be in agreement with that of Fermi liquid. Especially, the dispersion relation is linear. It is very different
from that found in the extremal Reissner-Nordstr$\ddot{o}$m (RN) black hole for massless fermion, which is obviously non-Fermi liquid
due to a nonlinear dispersion relation. However, for another scaling behavior of the height of $ImG_{11}$ at the
maximum, the scaling exponent is not one, which is at odds with that found in the RN black hole.

\end{abstract} \maketitle

\section {Introduction}

Motivated by the discovery of many exotic but important numerous correlated electron materials,
including the cuprate superconductors and other oxides,
lots of theoretical exploration on the quantum
phases and phase transitions of correlated electron systems are intensively processing
and varieties of phenomenological models were proposed in condensed matter theory.
However, a general theoretical framework characterizing such exotic states of matter remains a suspense.
Recently, many theoretical physicists have resorted to the AdS/CFT correspondence \cite{AdSCFT}
to offer a possible clue to yield the basic principle of such states of matter.
Indeed, we have found many of such exotic states of matter \cite{HongLiuNon-Fermi,FermiLiquidM}
using AdS/CFT correspondence.
At the same time, a proper condensed matter interpretation of these exotic states
has also been explored in recent studies \cite{Sachdev1,Sachdevet}.
Especially, in Ref.\cite{Sachdev1}, they have proposed that
the holographic non-Fermi liquid model in Ref.\cite{HongLiuNon-Fermi,HongLiuAdS2}
had realized an infinite-range limit of the fractionalized Fermi liquid \cite{FFL}.
Here, we must note that the two models share the same characteristics,
that is, nonzero ground state entropy density.

However, nonzero ground state entropy density in
Reissner-Nordstr$\ddot{o}$m black hole in Anti-deSitter spacetimes (RNAdS)
seems to be inconsistent with
the characteristics of the degenerate Fermi liquid at the zero temperature.
Therefore, a systematic exploration of the system in which
ground state entropy density vanishes will be important and valuable.
Such models have been proposed in Refs.\cite{ZeroGEG,ZeroGEO}.
In this paper, we will only focus on the model of Gubser and Rocha\footnote{We must point out that
in Ref.\cite{ZeroGEG}, the authors make the stability analysis of their supergravity solution and they find an instability. Therefore, the solution in the model of Gubser and Rocha is not the true ground state.
They presumed that it would involve some sort of charged scalar condensate or perhaps something completely different.}\cite{ZeroGEG}.
In this model, they construct a charged dilatonic black hole in $AdS_{5}$
in which the Lagrangian involves a gauge field and a neutral scalar.
In addition to the vanishing ground state entropy density mentioned above,
there is another important characteristic, that is, linear specific heat at low temperature.
It is also the characteristics of a Fermi gas.
Therefore, it is important and interesting to explore the properties of fermion response in this background.

Our paper is organized as follows.
In Sec. II, a brief review of the model of Gubser and Rocha is given.
Following Ref.\cite{HongLiuNon-Fermi}, we obtain
the Dirac equation of the probe fermions in this model in Sec. III.
Some properties are presented in Sec. IV.
Conclusions and discussions follow in Sec. VI.

\section {A charged dilatonic black hole geometry}

In this section, a brief review of the model of Gubser and Rocha is given as follows.
We will discuss the $AdS_{5}$ case and the $AdS_{4}$ case, respectively.

\subsection{$AdS_{5}$ case}

We begin with the following action:
\begin{eqnarray}
\label{DactionAdS5}
S=\frac{1}{2\kappa^{2}}\int d^{5}x \sqrt{-g}
\left[ R - 12(\nabla_{a}\alpha)^{2}
+ \frac{1}{L^{2}}( 8e^{2\alpha} + 4e^{-4\alpha})
-\frac{1}{4}e^{4\alpha} F^{ab}F_{ab}\right],
\end{eqnarray}
where $R$ is the Ricci scalar,
and $L$ is the AdS radius.
As is commonly known, $F_{ab}=\partial_{a}A_{b}-\partial_{b}A_{a}$.
The charged solution is\footnote{This black hole solution can be embedded in string theory.
For more discussions, we can refer to Ref.\cite{ZeroGEG}.}
\begin{eqnarray} \label{CdilatonBHSolutionAdS5}
ds^{2} &=& e^{2A} ( -h dt^{2} + d\vec{x}^{2} ) + \frac{e^{2B}}{h} dr^{2}, ~~~~~~~~~~ A_{a} dx^{a} = \Phi dt,
\nonumber\\ \label{s1AdS5}
A &=& \ln \frac{r}{L} + \frac{1}{3} \ln \left( 1 + \frac{Q^{2}}{r^{2}} \right), ~~~~~~~~
B = - \ln \frac{r}{L} - \frac{2}{3} \ln \left( 1 + \frac{Q^{2}}{r^{2}} \right),
\nonumber\\ \label{s2AdS5}
h &=& 1-\frac{\mu L^{2}}{(r^{2}+Q^{2})^{2}}, ~~~~~~
\Phi = \frac{ Q \sqrt{2 \mu} }{ r^{2}+Q^{2} } - \frac{ Q \sqrt{2 \mu} }{ r_{+}^{2}+Q^{2} }, ~~~~~~
\alpha = \frac{1}{6} \ln \left( 1 + \frac{Q^{2}}{r^{2}} \right).
\end{eqnarray}

The rescaled energy density, entropy density, and charge density can be easily expressed as follows
in the microcanonical ensemble:
\begin{eqnarray}
\label{EECdensity}
\hat{\epsilon} \equiv \frac{\kappa^{2}}{4\pi^{2}L^{3}}\epsilon=\frac{3\mu}{8\pi^{2}L^{6}}, ~~~~~~~~
\hat{s} \equiv \frac{\kappa^{2}}{4\pi^{2}L^{3}}s=\frac{\sqrt{\mu}r_{+}}{2\pi L^{5}}, ~~~~~~~~
\hat{\rho} \equiv \frac{\kappa^{2}}{4\pi^{2}L^{3}}\rho=\frac{Q\sqrt{2\mu}}{4\pi^{2} L^{5}}
.
\end{eqnarray}

With the help of Eq. (\ref{EECdensity}) and the condition $h(r_{+})=0$,
one can obtain the microcanonical equation of state:
\begin{eqnarray}
\label{MicroEofState}
\hat{\epsilon}=\frac{3}{2^{5/3}\pi^{2/3}} (\hat{s}^{2}+2\pi^{2}\hat{\rho}^{2})^{2/3}
.
\end{eqnarray}

Therefore, the temperature and chemical potential can be obtained as follows:
\begin{eqnarray}
\label{Temperature}
T=\left(\frac{\partial\hat{\epsilon}}{\partial \hat{s}}\right)_{\hat{\rho}}=\frac{r_{+}}{\pi L^{2}},~~~~
\Omega=\left(\frac{\partial\hat{\epsilon}}{\partial \hat{\rho}}\right)_{\hat{s}}=\frac{\sqrt{2}Q}{L^{2}}.
\end{eqnarray}

Then we can easily find
\begin{eqnarray}
\label{EntropyandT}
\hat{s}=\pi \sqrt{\frac{2 \hat{\epsilon}}{3}}T
\approx (\pi^{2} \hat{\rho})^{2/3}T
\approx \frac{\Omega^{2}}{4}T.
\end{eqnarray}
where the approximate equalities hold in the low-temperature limit.
In this paper, we will mainly focus on the zero-temperature limit.
From the expression of the temperature,
we find that the zero-temperature limit is obtained when $r_{+}=0$,
which implies $\mu L^{2}=Q^{4}$.
At the same time, we also notice that in this limit, the entropy vanishes,
which can have an important effect on the ground state.

In addition, from Eq. (\ref{EntropyandT}), we find that the rescaled specific heats
at constant charge density and constant chemical potential
\begin{eqnarray}
\label{SpecificH}
\hat{C}_{\hat{\rho}}=T \left( \frac{\partial \hat{s}}{\partial T} \right)_{\hat{\rho}},~~~
\hat{C}_{\hat{\Omega}}=T \left( \frac{\partial \hat{s}}{\partial T} \right)_{\hat{\Omega}},
\end{eqnarray}
are linear, which similar the case of a Fermi gas.

From the above discussions, we can see that there exists the similarity between
the low-temperature thermodynamics of the charged dilatonic black hole and the Fermi liquid.
Furthermore, in Ref.\cite{ZeroGEG}, they find the Fermi surface
for massless, charged bulk fermions at extremal limit by finding the normal modes.
It is plausible to claim the dual of the fermions of the charged dilatonic black hole
is a Fermi liquid. Therefore, in this paper, we will furthermore explore
the characteristics of the fermions response in this background.

\subsection{$AdS_{4}$ case}

The above case in $AdS_{5}$ can also be extended to the case in $AdS_{4}$ \cite{ZeroGEG,GubserEvolution}:
\begin{eqnarray}
\label{DactionAdS4}
S=\frac{1}{2\kappa^{2}}\int d^{4}x \sqrt{-g}
\left[ R - \frac{3}{2}(\nabla_{a}\alpha)^{2}
+ \frac{6}{L^{2}}\cosh \alpha
-\frac{1}{4}e^{\alpha} F^{ab}F_{ab}\right],
\end{eqnarray}

From the above action, one can obtain the charged solution
\begin{eqnarray} \label{CdilatonBHSolutionAdS4}
ds^{2} &=& e^{2A} ( -h dt^{2} + d\vec{x}^{2} ) + \frac{e^{2B}}{h} dr^{2}, ~~~~~~~~~~ A_{a} dx^{a} = \Phi dt,
\nonumber\\ \label{s1AdS4}
A &=& \ln \frac{r}{L} + \frac{3}{4} \ln \left( 1 + \frac{Q}{r} \right), ~~~~~~~~
B = - A,~~~~~~~~h = 1-\frac{\mu L^{2}}{(r+Q)^{3}},
\nonumber\\ \label{s2AdS4}
\Phi &=& \frac{\sqrt{3 Q \mu}}{r+Q} - \frac{\sqrt{3 Q }\mu^{\frac{1}{6}}}{L^{\frac{2}{3}}}, ~~~~~~
\alpha = \frac{1}{2} \ln \left( 1 + \frac{Q}{r} \right).
\end{eqnarray}

The extremal limit for $AdS_{4}$ is also obtained when $r_{+}=0$
(corresponding to $\mu L^{2}=Q^{3}$).
The case in $AdS_{4}$ shares the same characteristic
of low-temperature thermodynamics and has the Fermi surface.

\section {Dirac equation}

In this paper, we explore the fermion response in the charged dilatonic black hole
in $AdS_{5}$ and $AdS_{4}$, respectively.
First, we will briefly derive the Dirac equation in this background.
Subsequently, we discuss how to impose the boundary condition at the horizon
and read off the Green's function on the boundary.
For more details, we can also refer to Refs. \cite{HongLiuNon-Fermi,HongLiuAdS2}.

\subsection{The Dirac equation in a dilatonic black hole}

Considering the following bulk fermion action\footnote{Although
the background spacetime (\ref{CdilatonBHSolutionAdS5}) or (\ref{CdilatonBHSolutionAdS4})
comes from a UV-complete string theory, the fermion action (\ref{actionspinor}) is ad hoc
so that it is not necessary to derive it from type IIB supergravity.} \cite{HongLiuNon-Fermi}
\begin{eqnarray}
\label{actionspinor}
S_{D}=i\int d^{d+1}x \sqrt{-g}\overline{\zeta}\left(\Gamma^{a}\mathcal{D}_{a}-m\right)\zeta,
\end{eqnarray}
where $\Gamma^{a}$ is related to the usual
flat space gamma matrix by a factor of the vielbein,
$\Gamma^{a}=(e_{\mu})^{a}\Gamma^{\mu}$ and $\mathcal{D}_{a}=\partial_{a}+\frac{1}{4}(\omega_{\mu\nu})_{a}\Gamma^{\mu\nu}-iqA_{a}$
is the covariant derivative with $(\omega_{\mu\nu})_{a}$ the spin connection 1-forms.
The Dirac equation derived from the action $S_{D}$ is expressed as
\begin{eqnarray}
\label{DiracEquation1}
\Gamma^{a}\mathcal{D}_{a}\zeta-m\zeta=0.
\end{eqnarray}

Making a transformation
$\zeta=(-g g^{rr})^{-\frac{1}{4}}\mathcal{F}$ to remove the spin
connection and expanding $\mathcal{F}$ as $\mathcal{F}=F e^{-i\omega t +ik_{i}x^{i}}$
in Fourier space, the Dirac equation (\ref{DiracEquation1}) turns out to be
\begin{eqnarray}
\label{DiracEinFourier}
\sqrt{g^{rr}}\Gamma^{r}\partial_{r}F
-i(\omega+q A_{t})\sqrt{g^{tt}}\Gamma^{t}F
+i k \sqrt{g^{xx}}\Gamma^{x}F
-m F=0.
\end{eqnarray}
where due to rotational symmetry in the spatial directions,
we set $k_{x}=k$ and $k_{x}\neq 0,~i\neq x$ without losing generality.
Notice that Eq. (\ref{DiracEinFourier}) only depends on three Gamma matrices $\Gamma^{r},\Gamma^{t},\Gamma^{x}$.
So it is convenient to split the spinors $F$ into $F=(F_{1},F_{2})^{T}$ and
choose the following basis for our gamma matrices as in \cite{HongLiuAdS2}:
\begin{eqnarray}
\label{GammaMatrices}
 && \Gamma^{r} = \left( \begin{array}{cc}
-\sigma^3 \textbf{1}  & 0  \\
0 & -\sigma^3 \textbf{1}
\end{array} \right), \;\;
 \Gamma^{t} = \left( \begin{array}{cc}
 i \sigma^1 \textbf{1}  & 0  \\
0 & i \sigma^1 \textbf{1}
\end{array} \right),  \;\;
\Gamma^{x} = \left( \begin{array}{cc}
-\sigma^2 \textbf{1}  & 0  \\
0 & \sigma^2 \textbf{1}
\end{array} \right),
\qquad \ldots
\end{eqnarray}
where $\textbf{1}$ is an identity matrix of size $2^{\frac{d-3}{2}}$ for $d$ odd
(or size $2^{\frac{d-4}{2}}$ for $d$ even).
Because the resulting Green's functions are proportional to such an identity matrix,
we will suppress them in the following.
So, we have a new version of the Dirac equation as
\begin{eqnarray} \label{DiracEF}
\sqrt{g^{rr}}\partial_{r}\left( \begin{matrix} F_{1} \cr  F_{2} \end{matrix}\right)
+m\sigma^3\otimes\left( \begin{matrix} F_{1} \cr  F_{2} \end{matrix}\right)
=\sqrt{g^{tt}}(\omega+qA_{t})i\sigma^2\otimes\left( \begin{matrix} F_{1} \cr  F_{2} \end{matrix}\right)
\mp  k \sqrt{g^{xx}}\sigma^1 \otimes \left( \begin{matrix} F_{1} \cr  F_{2} \end{matrix}\right)
~.
\end{eqnarray}

Furthermore, according to eigenvalues of $\Gamma^{r}$,
we make such a decomposition $F_{\pm}=\frac{1}{2}(1\pm \Gamma^{r})F$. Then
\begin{eqnarray} \label{gammarDecompose}
F_{+}=\left( \begin{matrix} \mathcal{B}_{1} \cr  \mathcal{B}_{2} \end{matrix}\right),~~~~
F_{-}=\left( \begin{matrix} \mathcal{A}_{1} \cr  \mathcal{A}_{2} \end{matrix}\right),~~~~
with~~~~F_{\alpha} \equiv \left( \begin{matrix} \mathcal{A}_{\alpha} \cr  \mathcal{B}_{\alpha} \end{matrix}\right).
\end{eqnarray}

Under such decomposition, the Dirac equation (\ref{DiracEF}) can be rewritten as
\begin{eqnarray} \label{DiracEAB1}
(\sqrt{g^{rr}}\partial_{r}\pm m)\left( \begin{matrix} \mathcal{A}_{1} \cr  \mathcal{B}_{1} \end{matrix}\right)
=\pm(\omega+qA_{t})\sqrt{g^{tt}}\left( \begin{matrix} \mathcal{B}_{1} \cr  \mathcal{A}_{1} \end{matrix}\right)
-k \sqrt{g^{xx}} \left( \begin{matrix} \mathcal{B}_{1} \cr  \mathcal{A}_{1} \end{matrix}\right)
~,
\end{eqnarray}
\begin{eqnarray} \label{DiracEAB2}
(\sqrt{g^{rr}}\partial_{r}\pm m)\left( \begin{matrix} \mathcal{A}_{2} \cr  \mathcal{B}_{2} \end{matrix}\right)
=\pm(\omega+qA_{t})\sqrt{g^{tt}}\left( \begin{matrix} \mathcal{B}_{2} \cr  \mathcal{A}_{2} \end{matrix}\right)
+k \sqrt{g^{xx}} \left( \begin{matrix} \mathcal{B}_{2} \cr  \mathcal{A}_{2} \end{matrix}\right)
~.
\end{eqnarray}

Introducing the ratios $\xi_{\alpha}\equiv \frac{\mathcal{A}_{\alpha}}{\mathcal{B}_{\alpha}},\alpha=1,2$, and using the method developed in \cite{HongLiuUniversality,HongLiuSpinor,HongLiuAdS2},
one can package the Dirac equation (\ref{DiracEAB1}) and (\ref{DiracEAB2})
into the evolution equation of $\xi_{\alpha}$,
\begin{eqnarray} \label{DiracEF1}
(\sqrt{g^{rr}}\partial_{r}
+2m)\xi_{\alpha}
=\left[ \sqrt{g^{tt}}(\omega+q A_{t})+ (-1)^{\alpha} k \sqrt{g^{xx}}  \right]
+ \left[ \sqrt{g^{tt}}(\omega+q A_{t})- (-1)^{\alpha} k \sqrt{g^{xx}}  \right]\xi_{\alpha}^{2}
~,
\end{eqnarray}
which will be more convenient to impose the boundary conditions at the horizon
and read off the boundary Green functions.

\subsection{The boundary condition and the Green's function}

Near the boundary, a solution of the Dirac equation (\ref{DiracEF}) can be expressed as
\begin{eqnarray} \label{BoundaryBehaviour}
F_{\alpha} \buildrel{r \to \infty}\over {\approx} a_{\alpha}r^{m}\left( \begin{matrix} 0 \cr  1 \end{matrix}\right)
+b_{\alpha}r^{-m}\left( \begin{matrix} 1 \cr  0 \end{matrix}\right),
\qquad
\alpha = 1,2~.
\end{eqnarray}

If $b_{\alpha}\left( \begin{matrix} 1 \cr  0 \end{matrix}\right)$
and $a_{\alpha}\left( \begin{matrix} 0 \cr  1 \end{matrix}\right)$ are related by
$b_{\alpha}\left( \begin{matrix} 1 \cr  0 \end{matrix}\right)
=\mathcal{S}a_{\alpha}\left( \begin{matrix} 0 \cr  1 \end{matrix}\right)$,
then the boundary Green's functions $G$ is given by $G=-i \mathcal{S}\gamma^{0}$ \cite{HongLiuSpinor}.
Therefore
\begin{eqnarray} \label{GreenFBoundary}
G (\omega,k)= \lim_{r\rightarrow \infty} r^{2m}
\left( \begin{array}{cc}
\xi_{1}   & 0  \\
0  & \xi_{2} \end{array} \right)  \ ,
\end{eqnarray}

At the same time, the requirement that the solutions of Eqs. (\ref{DiracEAB1}) and (\ref{DiracEAB2})
at the horizon be in-falling implies
\begin{eqnarray} \label{GatTip}
\xi_{\alpha}\buildrel{r \to 0}\over =i,~~~~for~~\omega\neq 0.
\end{eqnarray}

Finally, we also note that $G_{11}$ and $G_{22}$ are related to each other as
$G_{22}(\omega,k)=G_{22}(\omega,-k)$. Therefore, we will focus solely on $G_{11}$ below.

\section{The properties of the Green's function}

\begin{figure}
\center{
\includegraphics[scale=0.9]{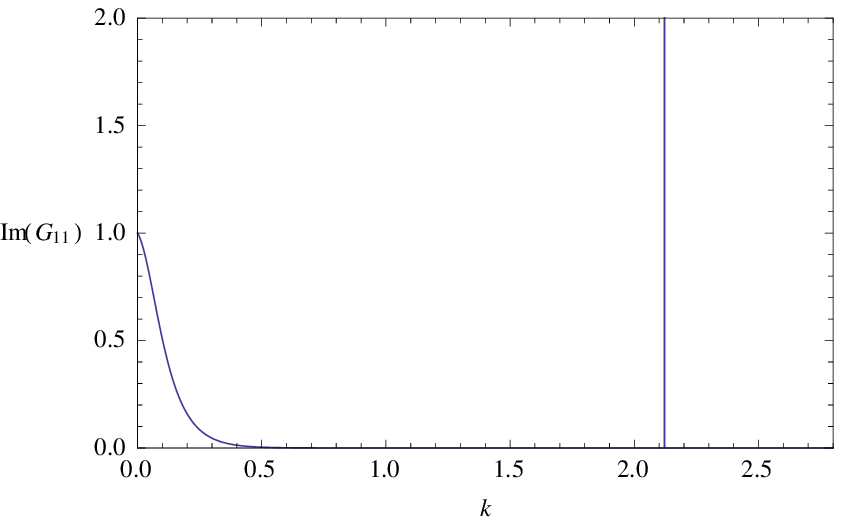}\hspace{1cm}
\includegraphics[scale=0.9]{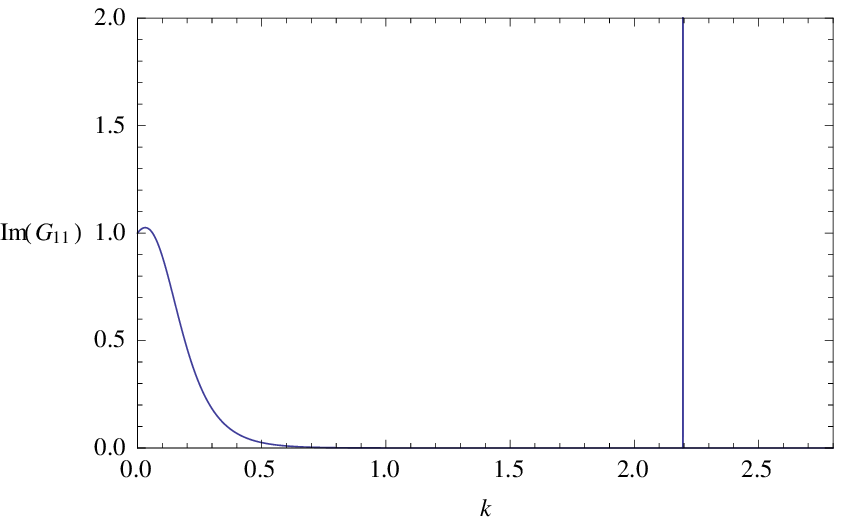}\\ \hspace{1cm}
\caption{\label{FermiM} The plot of $Im G_{11}(k)$ for $\omega=-10^{-9}$
(left plot for $AdS_{5}$ and right plot for $AdS_{4}$).
A sharp peak locates at $k_{F}\approx 2.12$ for $AdS_{5}$ and $k_{F}\approx 2.19$ for $AdS_{4}$.
Improving the accuracy, the Fermi momentum can be furthermore determined as $k_{F}\approx 2.121 305 342$
for $AdS_{5}$ and $k_{F}\approx 2.195 282 368 25$ for $AdS_{4}$.}}
\end{figure}
\begin{figure}
\center{
\includegraphics[scale=0.9]{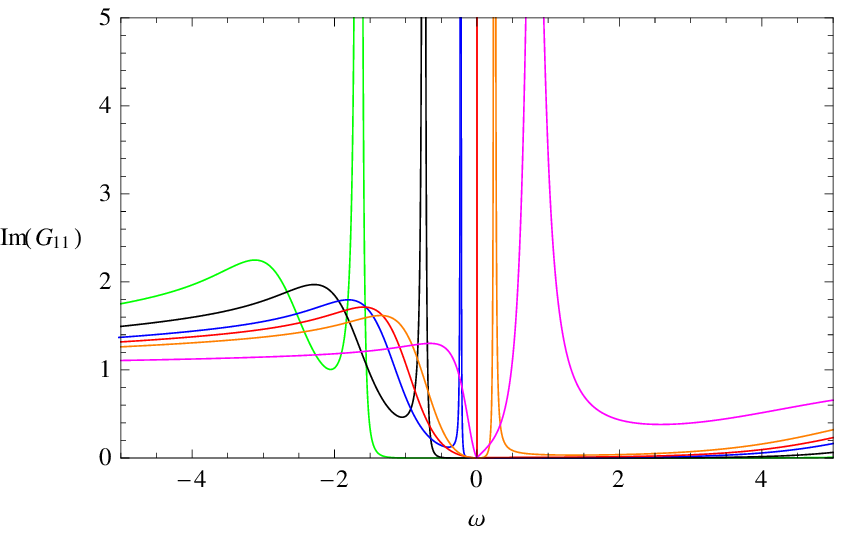}\hspace{1cm}
\includegraphics[scale=0.9]{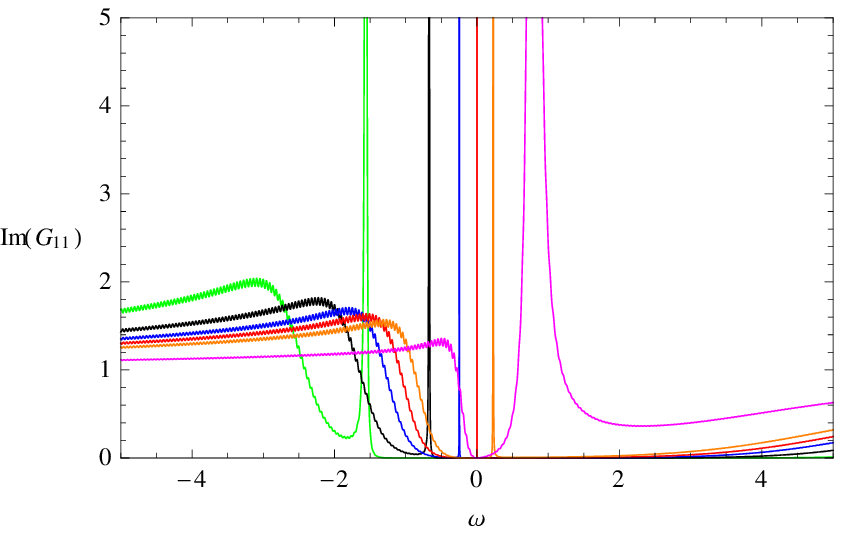}\\ \hspace{1cm}
\caption{\label{deltaF} The plot of $Im G_{11}(k)$ for various $k$
(left plot for $AdS_{5}$ and right plot for $AdS_{4}$).
They show that the quasiparticle peak approach a delta function at the Fermi momentum $k=k_{F}$.
Left (right) plot: green for $k=4$ ($k=4$), black for $k=3$ ($k=3$), blue for $k=2.4$ ($k=2.5$),
red for $k=2.12$ ($k=2.195$), orange for $k=1.8$ ($k=1.9$) and magenta for $k=0.8$ ($k=0.9$).
}}
\end{figure}

\begin{figure}
\center{
\includegraphics[scale=0.9]{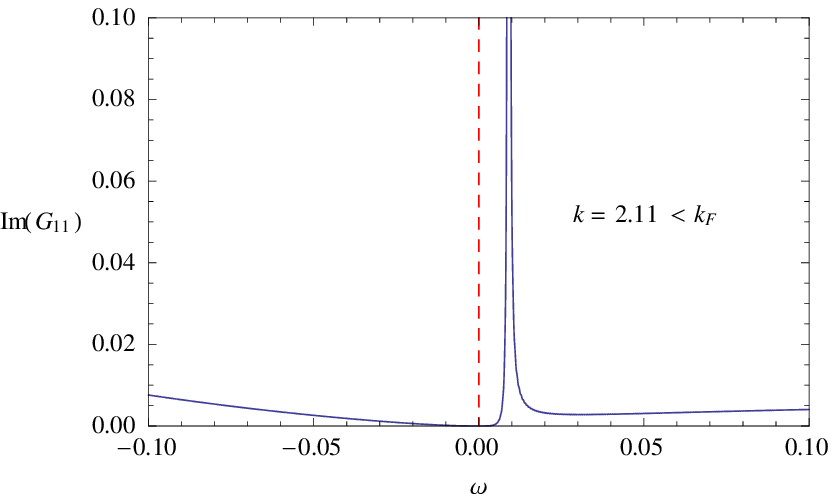}\hspace{1cm}
\includegraphics[scale=0.9]{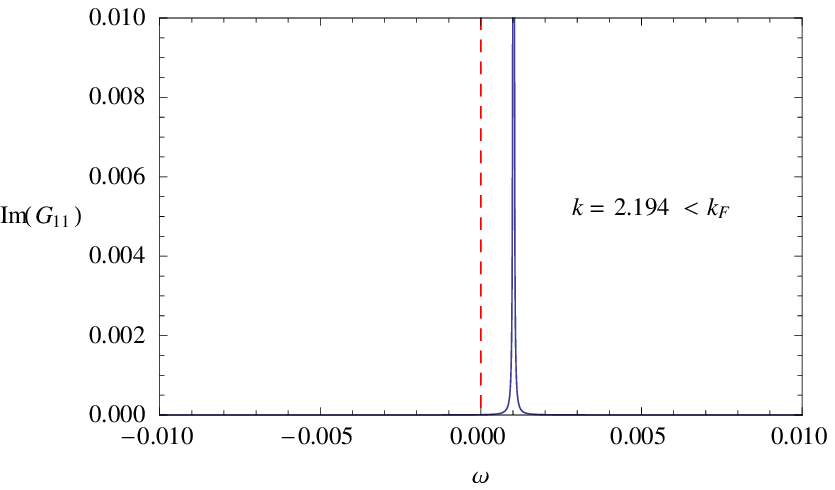}\hspace{1cm}
\includegraphics[scale=0.9]{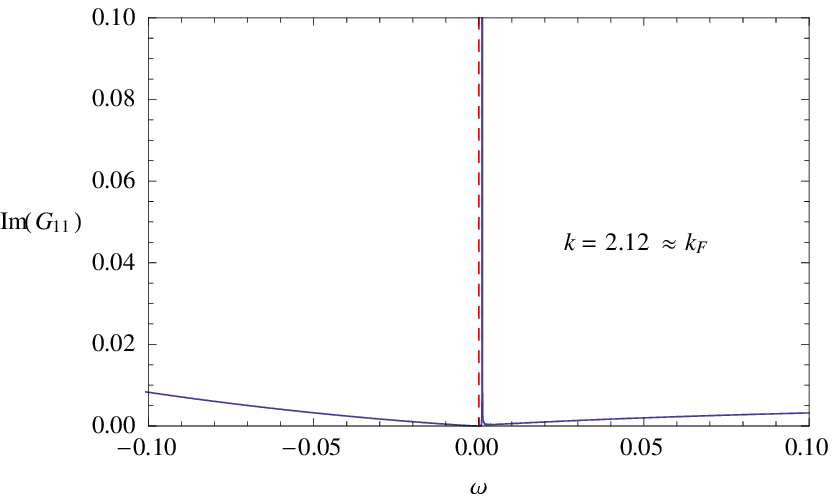}\hspace{1cm}
\includegraphics[scale=0.9]{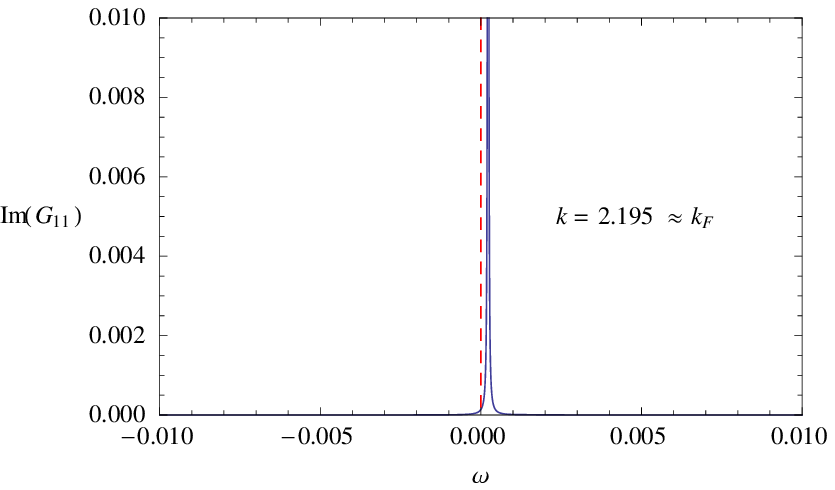}\hspace{1cm}
\includegraphics[scale=0.9]{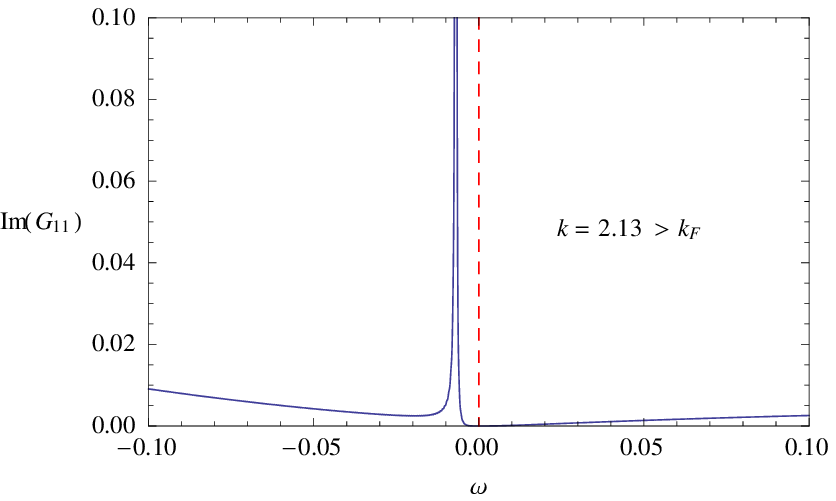}\hspace{1cm}
\includegraphics[scale=0.9]{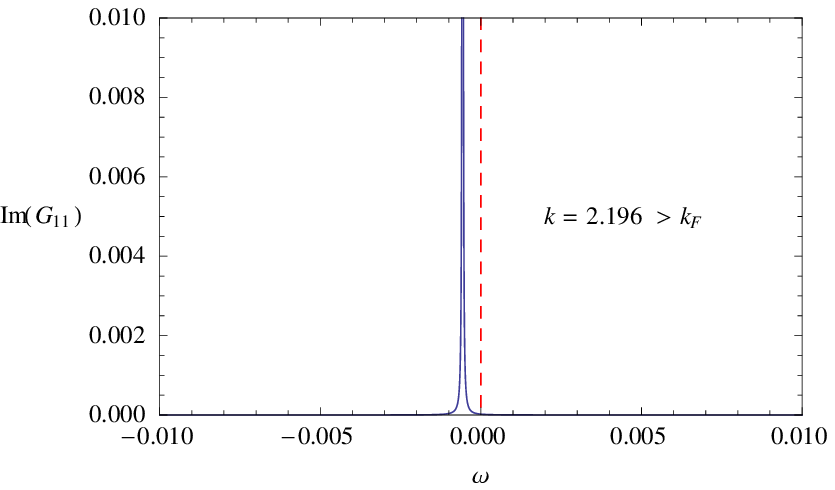}\\ \hspace{1cm}
\caption{\label{VanishingG} The plot of $Im G_{11}(k)$ for various $k$
(left plot for $AdS_{5}$ and right plot for $AdS_{4}$).
They show that independent of $k$, the Green's function vanishes at the Fermi energy.}}
\end{figure}
\begin{figure}
\center{
\includegraphics[scale=0.9]{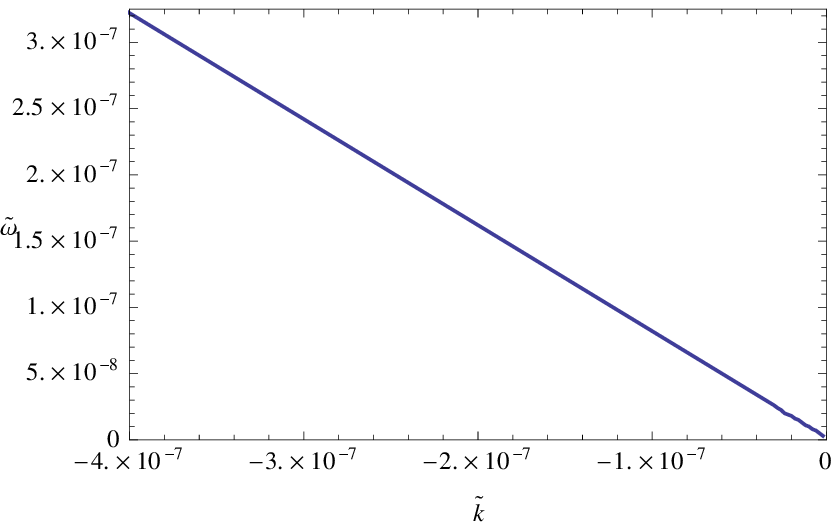}\hspace{1cm}
\includegraphics[scale=0.9]{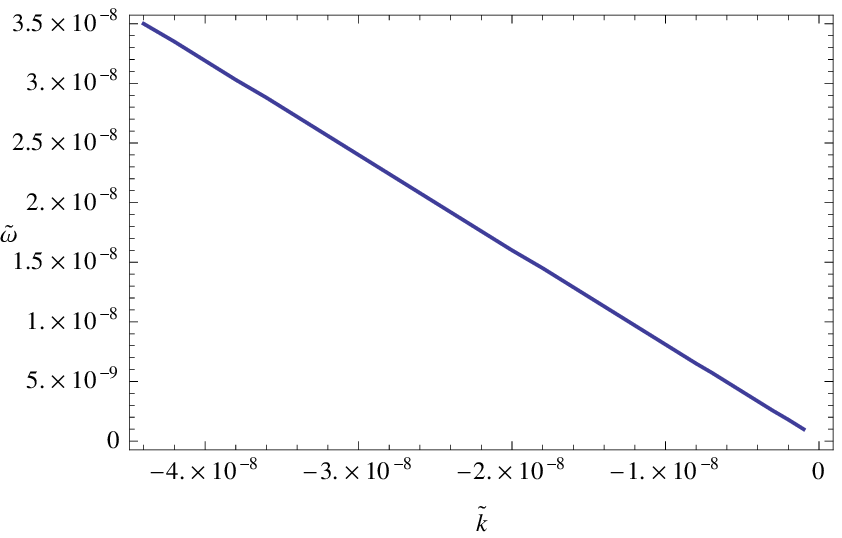}\\ \hspace{1cm}
\caption{\label{dispersion}The dispersion relation between $\tilde{k}$ and $\tilde{\omega}$ is linear
(left plot for $AdS_{5}$ and right plot for $AdS_{4}$).}}
\end{figure}
\begin{figure}
\center{
\includegraphics[scale=0.9]{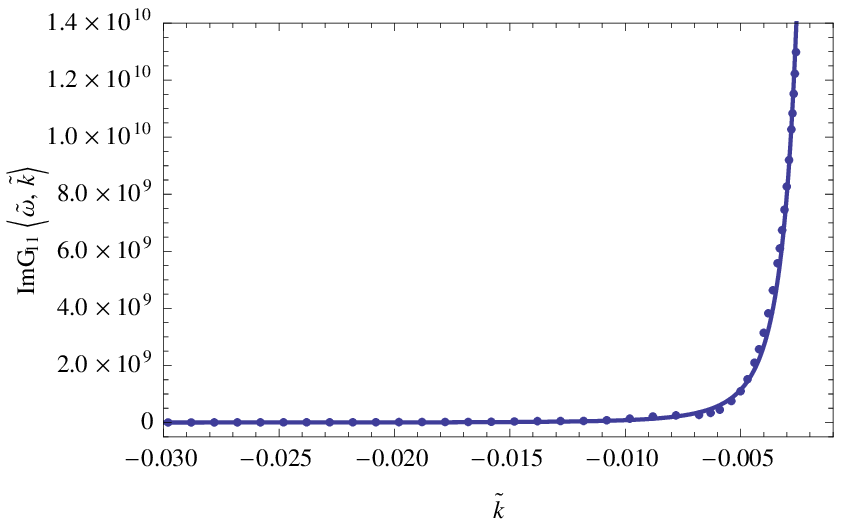}\hspace{1cm}
\includegraphics[scale=0.9]{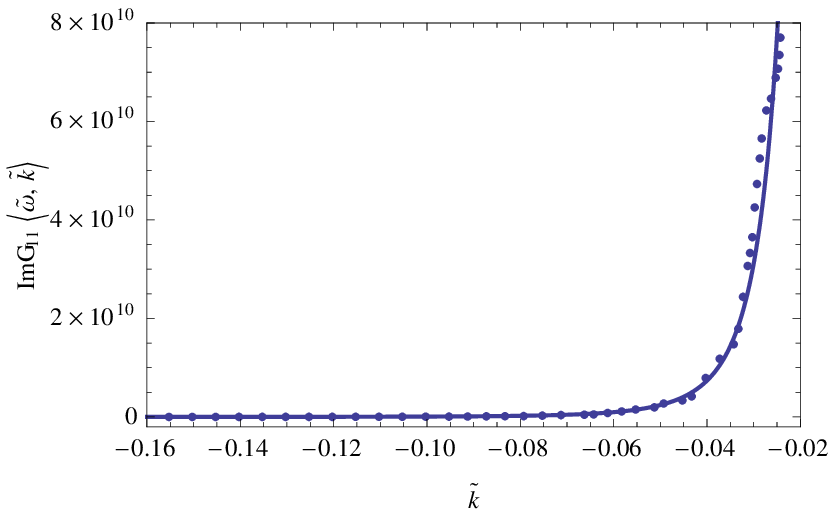}\\ \hspace{1cm}
\caption{\label{scalingpeak}The relation between the height of $ImG_{11}(\tilde{\omega},\tilde{k})$
 at the maximum and $\tilde{k}$ (left plot for $AdS_{5}$ and right plot for $AdS_{4}$).}}
\end{figure}

In this section, we solve Eq. (\ref{DiracEF1}) numerically with
the boundary conditions (\ref{GatTip}) to investigate the properties of the spectral function.
In Ref.\cite{ZeroGEG}, they find the Fermi surface by exploring the normal modes at $\omega=0$.
Here, we will follow the method developed in Ref.\cite{HongLiuNon-Fermi} to search for the Fermi surface.
We will consider the specific example as \cite{ZeroGEG}: $L=1,~m=0,~Q=1,~q=2$.

For $AdS_{5}$, when $\omega\rightarrow 0$,
a sharp quasiparticlelike peak occurs near $k_{F}= 2.121 305 34$ (left plot in Fig.(\ref{FermiM})).
While for $AdS_{4}$, such a peak located at $k_{F}\approx 2.195 282 368$ (right plot in Fig.(\ref{FermiM})).
It is consistent with the results found in Ref.\cite{ZeroGEG}.

Subsequently, we will explore the characteristics of the Green's function.
We find that:

\begin{enumerate}
\item At the Fermi momentum $k=k_{F}$, the quasiparticle peak is almost a delta function.
From Fig.\ref{deltaF}, we find that the peak becomes narrower and narrower when $k$ ($k>k_{F}$) decreases.
Finally, their heights approach infinity when $k$ approaches $k_{F}$.
However, when the Fermi surface is crossed, the peak become wide again.

\item The Green's function vanishes at the Fermi energy
(corresponding to $\omega=0$ in our convention),
which has nothing to do with $k$. It is shown in Fig.\ref{VanishingG}.

\item There exists a linear dispersion relation between small
$\tilde{k}$ and $\tilde{\omega}(\tilde{k})$ (Fig.\ref{dispersion}),
$i.e.$, $\tilde{\omega}(\tilde{k})\sim \tilde{k}$,
where $\tilde{k}\equiv k-k_{F}$ and $\tilde{\omega}(\tilde{k})$ is the location of the maximum of the peak.
It just corresponds to the form of a Fermi liquid.

\item Another important scaling behavior of the height of $ImG_{11}$ at the maximum is as follows:
\begin{eqnarray} \label{scalingHeight}
ImG_{11}(\tilde{\omega},\tilde{k})\sim \tilde{k}^{-\beta}
~,
\end{eqnarray}
where $\beta\simeq 3.8$ for $AdS_{5}$ and $\beta\simeq 5$
for $AdS_{4}$ (Fig.\ref{scalingpeak})\footnote{Here, we must point out that
due to the numerical instability, the momentum can not approach
the fermi momentum so close in pinning down the height of $ImG_{11}$
as pinning down $\tilde{\omega}$.}.
\end{enumerate}

The first three characteristics above, especially the third,
the linear dispersion relation, are like that of Fermi liquid.
It is very different from that found in the extremal RN black hole for massless fermions,
which is obviously non-Fermi liquid due to the nonlinear dispersion relation \cite{HongLiuNon-Fermi}.
However, another important scaling behavior of the height of $ImG_{11}$ at the maximum
is obviously at odds with that found in Ref.\cite{HongLiuNon-Fermi},
in which the scaling exponent $\beta\simeq 1$.
It is a peculiar property of Green's functions of the fermions in this system.
Therefore, it is interesting and important to furthermore explore the properties of holographic fermions
in the extremal charged dilatonic black hole.

\section{Conclusions and discussion}

We have studied the main features of the fermions in the charged dilatonic black hole
for zero-temperature limit and massless fermions by AdS/CFT correspondence.
Many features of the fermions' Green's function, especially the linear dispersion relation,
and the low-temperature thermodynamics studied in Ref.\cite{ZeroGEG} are plausible that the dual is Fermi liquid.
However, for another scaling behavior of the height of $ImG_{11}$ at the maximum,
the exponent $\beta$ is not one, indicating the case may be not so simple.
Nevertheless, due to the peculiar properties of this model,
it will be valuable to furthermore explore the characteristics of the fermions response in this background.
In addition, it is also important to relate this
model to the state of condensed matter.

\begin{acknowledgments}

I would like to thank Professor Yongge Ma, for his encouragement.
I am also grateful to Professor Yi Ling and Dr. Wei-Jie Li for their useful discussions.
In addition, I also thank F. D. Rocha for his valuable comments on the true ground state.
This work is partly supported by NSFC(No.10975017) and the Fundamental Research
Funds for the central Universities.

\end{acknowledgments}

\end{document}